\documentclass[aip, reprint]{revtex4-1}

\usepackage{graphicx}% Include figure files
\usepackage{dcolumn}% Align table columns on decimal point
\usepackage{bm}% bold math
\usepackage[utf8]{inputenc}
\usepackage[T1]{fontenc}
\usepackage{mathptmx}
\usepackage{etoolbox}
\usepackage{gensymb}
\usepackage{xcolor}
\usepackage{amssymb}
\usepackage{amsmath}

\def\permm{\mathrm{m^{-2}}}
\def\permmm{\mathrm{m^{-3}}}
\def\persec{\mathrm{s^{-1}}}

\def\2D{\mathrm{2D}}
\def\3D{\mathrm{3D}}

\def\Wpermm{\mathrm{W\cdot m^{-2}}}

\def\ve{v_e}
\def\ne{n_e}
\def\CMR{Q/M}

\def\AmCa{$(\mathrm{NH}_4)_2\mathrm{CO}_3~$}

\def\h2o{$\mathrm{H}_2\mathrm{O}$}
\def\co2{$\mathrm{CO}_2$}
\def\c2h4{$\mathrm{C_2H_4}$}

\def\o2{$\mathrm{O}_2$}
\def\N2{$\mathrm{N}_2$}
\def\H2{$\mathrm{H}_2$}

\def\h+{$\mathrm{H}^+$}
\def\Au+{$\mathrm{Au}^+$}
\def\e-{$\mathrm{e}^-$}

\def\psat{p_{\mathrm{sat.}}}

\begin{document}

%\preprint{}

\title{
    \begin{center}
        Carbon encapsulation of levitated Au nanoparticles
    \end{center}} %Title of paper



\author{Joyce E. Coppock}
    \email{jec@terpmail.umd.edu}
    \affiliation{University of Maryland, College Park, MD, 20742, USA}
    \affiliation{Laboratory for Physical Sciences, 8050 Greenmead Dr., College Park, MD, 20740, USA}
\author{Sunghyun Kim}
    \email{shdak6990@gmail.com}
    \affiliation{University of Maryland, College Park, MD, 20742, USA}
    \affiliation{Laboratory for Physical Sciences, 8050 Greenmead Dr., College Park, MD, 20740, USA}

 \author{B. E. Kane}
    \email{bekane@umd.edu}
    \affiliation{Laboratory for Physical Sciences, 8050 Greenmead Dr., College Park, MD, 20740, USA}
    \affiliation{Joint Quantum Institute, University of Maryland, College Park, MD, 20742, USA}


\date{29 June 2026} %changed from "\today" to set date for Arxiv submission purposes

\begin{abstract}
We investigate the formation of a barrier to evaporation that develops when levitated nanoscale Au nanoparticles are exposed to pulses of 532 nm laser radiation in a high vacuum (pressure $p=10^{-8}-10^{-7}$ Torr) environment.  Our data are derived from precision measurements of the charge to mass ratio ($\CMR$) of $\sim$200 nm diameter Au particles confined in a quadrupole ion trap.  We characterize the development of the barrier over time as the particle is repeatedly heated with laser pulses and determine the impact of variations of the interval between pulses and of exposure to several gases added to the vacuum chamber. We observe a slow increase in the mass of particles upon prolonged exposure to the vacuum, which we attribute to the growth of a barrier layer.  For particles that have acquired a barrier during exposure to CO, we observe a rapid decrease in their mass upon subsequent exposure to \o2.  These findings are consistent with the growth and subsequent oxidation of a graphene layer on the Au that forms the barrier to evaporation.  However, we have not found that the rate of formation of the barrier depends on the pressure of carbon-containing gases (CO, \c2h4, \co2) we have added to the chamber. We hypothesize that a rare surface state on the solid Au particle catalyzes the reaction that introduces C to the particle. Repeated laser pulse heating is necessary---either to enable diffusion away from this state or to create fresh states that allow continued C  uptake---to facilitate the growth of the surface graphene layer.
\end{abstract}

\pacs{}% insert suggested PACS numbers in braces on next line

\maketitle %\maketitle must follow title, authors, abstract and \pacs


\section{Introduction}
\label{sec:intro}

The ability to levitate nanoscale materials and probe their properties in a controlled environment has created new opportunities for the exploration of the chemistry of single nanoscale particles and their surfaces. For example, these techniques enable measurement of materials at very high temperatures\cite{Friese2025}, including measurements of C oxidation \cite{Rodriguez2021} and sublimation\cite{Long2020} at temperatures $T\geq$1200K.  In part because Au nanoparticles with well-specified size are readily available, levitated Au has received significant attention, and the effects of the surface plasmon resonance (SPR)  are of particular interest \cite{Hoffman2023}, since levitated particles are commonly probed at wavelength $\lambda=$ 532 nm, where effects of SPR are most pronounced.

Our previous work has focused on the melting\cite{Coppock2021} and undercooling\cite{Coppock2022} of levitated Au nanoparticles subjected to 532 nm laser pulses.  While these investigations were performed in a system capable of reaching high vacuum (pressure $p\leq10^{-7}$ Torr), we obtained reproducible data only in a $10^{-5}-10^{-6}$ Torr oxygen ambient.  We attributed this phenomenon to adventitious C on the Au surface that tends to accumulate in an imperfect vacuum unless sufficient \o2 is present for oxidation to remove the contaminants.  Adventitious C has been implicated in the contamination of other surfaces, such as diamond, in vacuum environments; such contamination can sometimes be removed by \o2.\cite{Parthasarathy2024}  Our data suggested that C on the Au surface creates an impermeable barrier layer that tends to suppress Au evaporation. 

Because of their relevance to the growth of graphene and carbon nanotubes, gas reactions at Au surfaces that yield solid C have been well investigated\cite{Oznuluer2011,Terasawa2019,Lu2022}. Reactions of CO on Au surfaces, photocatalyzed by 532 nm radiation,\cite{Hung2008,Hung2010} may be particularly relevant to interpreting our previous results in light of the fact that CO is a common residual gas in vacuum chambers at high vacuum.  Complete carbon encapsulation of nanoscale Au has been reported\cite{Ugarte1993,Wu2012,Liu2018}, indicating that surface C reactions can proceed until the Au surface is almost entirely surrounded by graphene or graphitic material.

We describe below our experimental work to elucidate the mechanism for evaporation suppression and barrier formation on levitated Au samples in vacuum and in various gas ambients.  First, we discuss our procedures for preparation, collection, and cleaning of the charged 200-250 nm Au nanoparticles that we use in our experiments.  We also describe the procedures we use to expose levitated particles to different gases, and we identify the residual gases present in our imperfect vacuum that are possibly relevant to our observations.

All of our data are deduced from precision measurements of the charge to mass ratio ($\CMR$) and the mass ($M$) of the levitated particle as it is subjected to continuous wave (CW) and pulsed 532 nm laser irradiation. Laser pulses bring the particle  $T$ into a regime where the evaporation rate can be determined from the mass change, $\Delta M$, caused by a pulse of known width $w$. Using data from repeated sequences of pulses that heat the particle to the neighborhood of the Au melting point, we are able to measure the suppression of evaporation. We characterize this suppression by a parameter $\sigma$, where $\sigma=1$ is a clean surface and $\sigma\rightarrow0$ is a surface entirely impermeable  to Au evaporation.

We measure the effect on $M$ and $\sigma$ of varying the interval $\Delta t$ between heating pulses and of introducing various gases into the vacuum chamber.  We also measure the effect of exposing the surface to heating in an \o2 ambient ($p=2-3\times10^{-5}$ Torr) after it has been contaminated.  Our most notable  findings are twofold.  First, for particles left in an imperfect vacuum for long periods, we observe a mass $gain$ that approaches that of a single layer of graphene on the particle surface.  Conversely, when a particle with surface evaporation suppressed ($\sigma<1$) by pulses in CO ambient ($p=6-7\times10^{-6}$ Torr) is subsequently exposed to \o2, we observe a mass $loss$ of comparable magnitude.  Immediately after this mass loss, the particle is free of contamination, and $\sigma\approx1$.

While this evidence is certainly suggestive that a graphene barrier is responsible for the suppression of evaporation that we observe, we have not been able to correlate the rate of growth of this barrier to the pressure of any of the C-containing gases that we have added to the chamber (CO, \c2h4, and \co2). This null result leads us to posit that a rare surface state, possibly related to the edges and corners of the Au when it is a solidified crystal, is a necessary site for a reaction that converts a volatile carbonaceous gas to a C atom in the Au. The presence of the C product at the site hinders further reactions until the C diffuses away during high temperature pulses or the site is reset by melting and recrystallization of the Au.  High temperatures also enable the C to migrate and attach to a surface graphene layer that forms the barrier to Au evaporation.

\begin{figure}
%\begin{center}
\includegraphics[scale=1.0,draft=false]{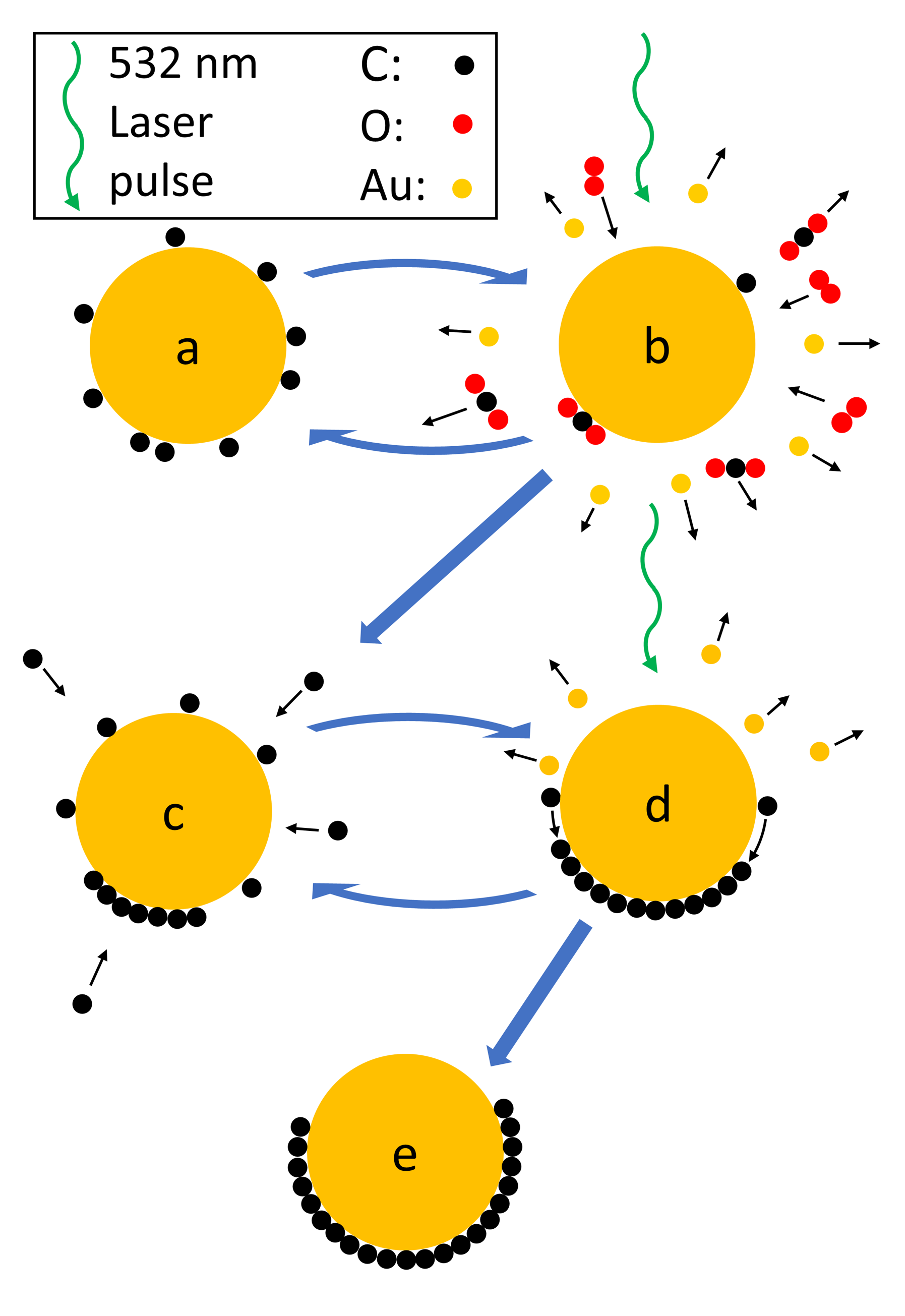} 
%\end{center}
\caption{
Proposed mechanism for preparation of a clean surface on an Au nanoparticle and its subsequent encapsulation in C. (a-b) Carbonaceous materials originally on the nanoparticle are removed by repeatedly pulsing the sample temperature above the melting point of Au in \o2 ambient using a 532 nm laser. After these cleaning steps, Au can freely evaporate from the nanoparticle surface at high temperatures. In the absence of \o2, however, adventitious carbon (probably a component of a volatile gas, but illustrated here as a single C atom) can accumulate on the surface and begin to form 2D layers on the surface after repeated exposure to heat pulses (c-d).  The 2D carbon forms a barrier to Au evaporation (e) leading to a suppressed evaporation rate and reduced mass loss of the particle during high temperature pulses.
}
\label{Illustration}
\end{figure}

\section{Methods}
\subsection{Au nanoparticle preparation and cleaning}
\label{sec:preparation}

As we have described previously\cite{Kane2010,Coppock2021}, nanoparticles levitated in our apparatus originate from liquid suspensions that are injected into the vacuum system via electrospray ionization.  We use commercially available Au nanoparticles of nominal initial diameter 250 nm\cite{Cytodiagnostics} ($M=1.6\times10^{-16}$ kg) that are packed in pure deionized water.  The particles are rinsed and the water replaced with a suspension medium suitable for electrospray, as described in an earlier paper\cite{Coppock2021}.  The suspension medium is prepared by diluting \AmCa to a concentration of 2 mM using a solution of 1 part isopropyl alcohol to 3 parts deionized water.
%The electrospray plume is directed through an aperture of diameter 100 um leading into a vacuum chamber to allow a particle to be collected in an ion trap inside the chamber (P$\sim$500 mTorr).  
After capture, the particle is transferred to a second trap in a high-vacuum chamber using a dual trap linear shift mechanism\cite{Coppock2017}.  
%The chamber is backfilled with \o2 to a pressure around 30 mTorr to stabilize the particle during transfer.  

After transfer, experiments are conducted at high vacuum pressures, usually in the range of 10$^{-8}$ to 10$^{-5}$ Torr.  The chamber is continuously evacuated with a turbo pump backed by an oil-lubricated mechanical pump.  Prior to the series of experiments described in this paper, the vacuum system was baked to $\sim$100$\degree$ C, resulting in an ultimate vacuum pressure of $p=(1-2)\times10^{-8}$ Torr.  The most prevalent gas species after bakeout, as measured by a residual gas analyzer (RGA), are shown in Table \ref{TableRGATurbo}.  The most likely sources of carbon in the system are CO and CO$_2$.  The strongest peak in the RGA spectrum (aside from $\mathrm{H}_2\mathrm{O}$) 
is at 28 amu, which is likely due primarily to CO.  N$_2$ is a less likely candidate because we do not detect Ar (a typical component of an air leak) at all.  \c2h4 is unlikely to contribute significantly to the 28 amu peak, because peaks at 26 and 27 amu, corresponding to fragments of the parent gas, were not detected. The 16 amu peak could be due to atomic O or CH$_4$; however, O is probably a stronger contributor, since peaks at 14 and 13 amu (which would be likely fragments of CH$_4$) are not detected.  The 36 amu line is likely either hydrated ammonium cations from electrospraying of the suspension medium or water dimer cations\cite{Song2025}.

Chamber pressures are measured by two hot-filament ionization gauges: one located in a gauge port near the chamber and a supplementary one near the turbo pump.  Pressures cited in this paper are measured near the chamber unless otherwise indicated.  For certain experiments, small amounts of bottled gases\cite{GasBottles} are added to the vacuum chamber via high vacuum leak valves.  When \o2 is added, the total chamber pressure is maintained in the range of $p=(2-3)\times10^{-5}$ Torr; this configuration will be referred to as ``\o2 ambient.''  In the \o2 ambient configuration, the pressure gauge near the chamber is turned off (in order to avoid excessive exposure of the filament to \o2) and the pressure is monitored by the gauge near the turbo pump.  When gases other than \o2 are added, the chamber pressure is maintained in the range $p=(6-7)\times10^{-6}$ Torr unless stated otherwise; this configuration will be referred to as a ``gas ambient.''  For experiments where no gases are added, chamber pressure varies from $p=(2-5)\times10^{-8}$ Torr.  This configuration will be referred to as ``vacuum.''  

%Since the chamber is frequently vented to medium vacuum pressures, the population of gas species is not as well-controlled as it would be in a system that is baked out and kept at high vacuum constantly.  

While the particle is levitated in the high-vacuum chamber at $P\leq10^{-5}$ Torr, particle oscillation amplitudes are stabilized by parametric feedback\cite{Kane2026}.  The particle is constantly illuminated with a 300 $\mu$W CW laser (with power density in the region of the particle approximately 1$\times10^{3}~\Wpermm$) as an optical probe.  Oscillation frequencies (from which $\CMR$ and $M$ are deduced using formulas in Section \ref{sec:acquisition}) are measured with precision $\sim$10$^{-5}$ (Ref. \citenum{Kane2026}).
During initial pumpdown of the chamber, $\CMR$ is found to decrease in discrete, uniform increments, which are identified as losses of individual electronic charges, of magnitude $|e|$.  (All of our samples are positively charged.)  The magnitude of these increments are used to determine the net charge $\ne=Q/e$, as described in Sec.~\ref{sec:acquisition}.  The particle discharges at a fast rate upon initial pumpdown; we attribute this to the removal of loosely attached charged residue, which probably becomes attached to the particles when they are in the liquid suspension.  

If the particle is left in high vacuum at a low probe laser power, electric discharge will continue at a gradually slowing rate for days.   In order to expedite the discharge process, the particle is subjected to a standard ``heat treatment” or cleaning process, which is performed in an \o2 ambient because oxygen appears to facilitate surface cleaning.  In the first step of the heat treatment process, the particle is subjected to a laser pulse with $w=$20 ms and maximum power density $\sim1\times10^{5}~ \Wpermm$ in the region of the particle.  A drop in the measured $\CMR$ value occurs that corresponds to the loss of a fraction of the particle’s total charge.  Figs.~\ref{Illustration}a and \ref{Illustration}b depict charged material being detached from an Au particle upon heating. The charge loss varies between samples, typically from about 0.5\% to 25\% of the n
et charge.  In occasional cases ($\sim$20\% of recent samples), this initial pulse causes either particle loss or discharging so severe that the particle becomes unsuitable for study in the system ($\CMR<0.5$ C/kg).
%either because it loses so much charge that it can no longer be trapped, or because the detachment of charged material imparts enough momentum to expel it from the trap.  

Our subsequent experiments require discharge events to be relatively rare, so that changes in $\CMR$ during laser pulses can be attributed to changes in M with high likelihood.  In order to verify that this condition has been reached, a series of 10 identical laser pulses are applied to the particle, with $w=$70 ms and the same power as the first pulse.  The width is chosen to be long compared to the time taken for the particle to reach its equilibrium temperature\cite{Coppock2021}. 
%During each pulse, about 0.1% of the particle's mass is evaporated.  
$\CMR$ is measured before and after each pulse.  If no charge is lost at the time of each pulse, the same $\CMR$ increase will be observed on all pulses.  In practice, discharge does not fully cease during this process, but the rate can be confirmed to be low enough that an acceptably small number of data points will have errors caused by charge loss.
Further discussion of outliers caused by discharge can be found in Sec.~\ref{sec:qualitative}.
 Mass and net charge after the heat treatment process (just before the beginning of experiments) are shown in Table \ref{TableMassCharge}, in the ``Initial $M$” and ``Initial $\ne$” columns, respectively.

\begin{table}
\begin{tabular}{|c|c||c|c|c|}   \hline\hline
Atomic mass&Likely species&Partial pressure \\ 
(amu)& &(Torr) \\
\hline
2& H$_2$ &2.1$\times10^{-9}$ \\
16& O, CH$_4$ &6.0$\times10^{-10}$ \\
17& OH$^-$ &1.4$\times10^{-9}$ \\
18& H$_2$O &4.0$\times10^{-9}$ \\
28& CO, N$_2$, C$_2$H$_4$ &2.8$\times10^{-9}$ \\
32& O$_2$ &4.7$\times10^{-10}$ \\
36& (H$_2$O)$_2^+$, NH$_4^+\cdot$ H$_2$O &9.8$\times10^{-10}$ \\
44& CO$_2$ &1.5$\times10^{-9}$ \\ \hline\hline
\end{tabular}
\caption{RGA data showing partial pressures for the most populous gas species after baking out the vacuum system.  Total pressure: $p=1.4\times10^{-8}$ Torr.}
\label{TableRGATurbo}
\end{table}

\begin{table}
\begin{tabular}{|c||c|c||c|c|}   \hline\hline
Sample index&Initial $M$&Initial $\ne$&Final $M$&Final $\ne$\\ 
 &(kg)& &(kg)& \\
\hline
250416& 1.10$\times10^{-16}$ & 1004 & 6.34$\times10^{-18}$ & 149 \\
(26)0108& 1.09$\times10^{-16}$ & 1461 & 6.91$\times10^{-17}$ & 1371 \\
(26)0221& 1.13$\times10^{-16}$ & 925 & 6.56$\times10^{-17}$ & 892 \\
(26)0310& 1.09$\times10^{-16}$ & 1237 & 5.90$\times10^{-17}$ & 931 \\
(26)0409& 1.43$\times10^{-16}$ & 1831 & 5.76$\times10^{-17}$ & 1634\\
(26)0507& 1.07$\times10^{-16}$ & 1568 & 2.79$\times10^{-17}$ & 1270 \\
(26)0519& 9.42$\times10^{-17}$ & 1363 & 2.77$\times10^{-17}$ & 1230 \\ \hline\hline
\end{tabular}
\caption{Table of mass $M$ and net charge $\ne$ for selected nanoparticle samples.  Uncertainty in the tabulated values of $\ne$ (and consequently $M$) are typically $\leq$3\% (see Ref.~\citenum{Kane2026}). Sample indices are based on the date (YYMMDD) when data taking began.  In figures, samples from 2026 will be referred to by the last four digits in their index.  Initial mass and charge were measured after heat treatment cleaning procedure, just before first data set in \o2.  Final mass and charge were measured after last plotted data set for each particle.}
\label{TableMassCharge}
\end{table}

\subsection{Data acquisition}
\label{sec:acquisition}

\begin{figure}
%\begin{center}
\includegraphics[scale=1.0,draft=false]{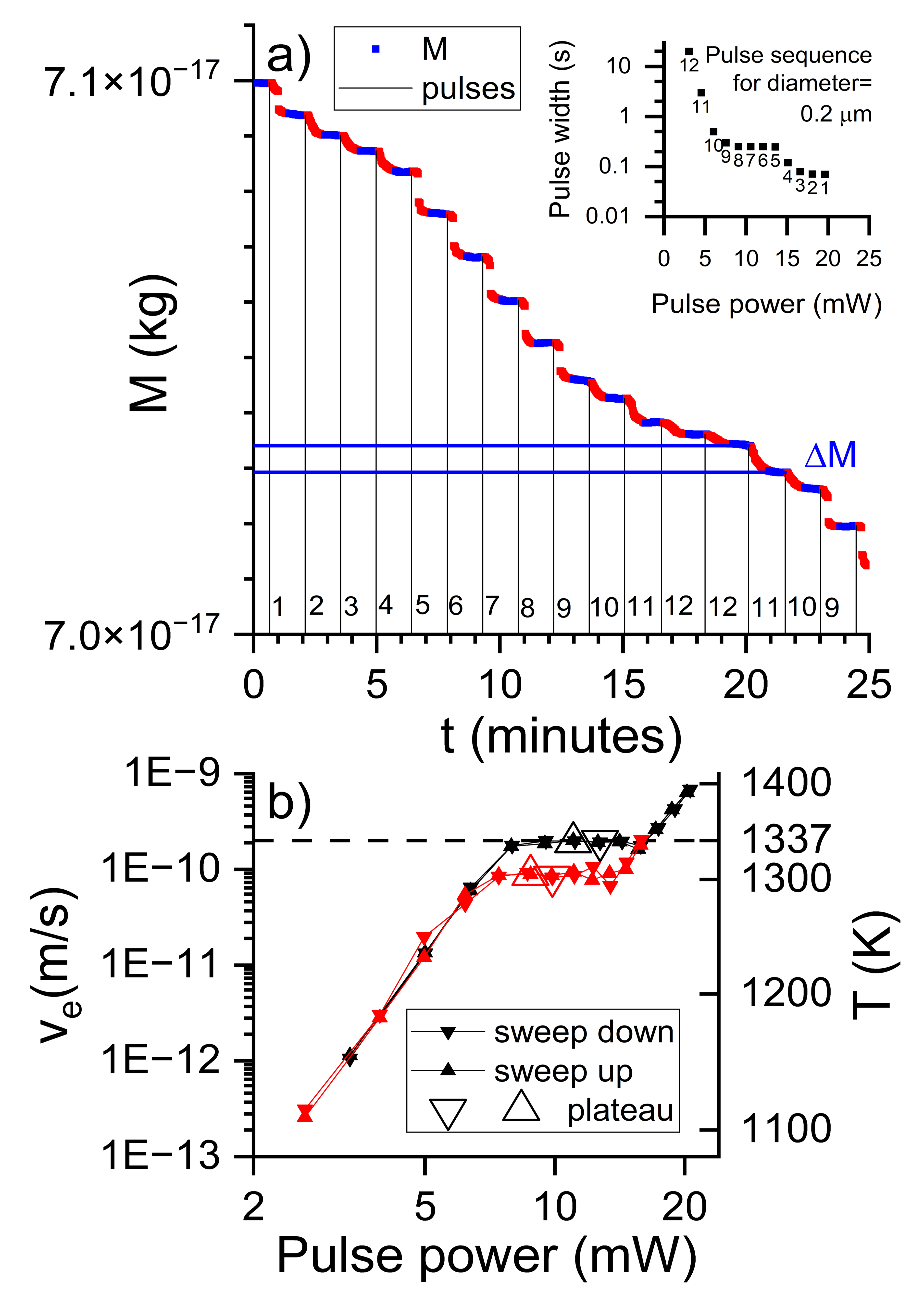} 
%\end{center}
\caption{
(a) Mass, $M$, is determined from measurements of $\CMR$ and the value of $Q=e~n_e$ inferred from discharge steps. In this data, the illumination laser is pulsed every $\approxeq$90 seconds.  The pulse power follows a triangle wave pattern, with pulse widths $w$ for a particle of diameter 200 nm shown in the inset.  The measured $\Delta M$ and $w$ are used in Eq. \ref{erosion velocity} to determine the erosion velocity, $\ve$. (b) $\ve$, plotted as a function of laser pulse power, showing the plateau that is characteristic of Au melting. In our experiments on clean samples pulsed in a \o2 ambient, the plateau occurs near the $\ve$ value expected for the melting temperature of Au=1337 K (black data).  The plateau is suppressed, however, if samples are subjected to repeated pulses in vacuum (red data).  Labels are used to designate whether the data was obtained using upward ($\blacktriangle$) or downward ($\blacktriangledown$) sweeps in $T$. Large triangles are an estimate of the characteristic value of $\ve$ on the plateau and are the values from which $\sigma$ is determined in subsequent data.
}
\label{IntroPlot}
\end{figure}

All of our measurements rely on the fact that $\CMR$ of trapped charged particles can be measured to high precision \cite{Kane2026} and that the number of charges on the particle can be inferred from observation of single charging or discharging steps in $\CMR$:

\begin{equation} \label{Q/M}
M=\frac{e ~\ne}{\CMR}.
\end{equation}
In practice, we do not observe spontaneous increases in the net charge, only decreases.
Our data is taken using a sequence of laser pulses (Fig.~\ref{IntroPlot}a) sufficiently powerful to evaporate some of the Au in the nanoparticle, separated by a interval when the CW power of the laser has a small to negligible effect on the particle mass. The interval between pulses (blue data) is used to acquire data to determine the mass. The width of the pulses, $w$, is much longer than the time it takes for the particle to reach a constant temperature \cite{Coppock2022}, so the characteristic evaporation rate (or erosion velocity) $\ve$ can be derived from:

\begin{equation} \label{erosion velocity}
\ve\equiv \frac{dr}{dt}=\frac{\Delta M}{w}\times\frac{1}{4 \pi r^{2} \rho},
\end{equation}
where $r$ is the radius of the particle, which we assume to be spherical. $\rho$ is the Au density, which for the solid near the melting point ($T=$ 1337.3 K) is 18,300 kg$\cdot\permmm$ (Ref. \citenum{Pamato2018}). The theoretical value of $\ve$ can in turn be determined from the vapor pressure $\psat$of Au\cite{Alcock1984} and the mass of single Au atoms, $m_{\mathrm{Au}}$:

\begin{equation} \label{psat}
\ve=  -\frac{\psat}{\rho} \sqrt \frac{m_{\mathrm{Au}}}{2 \pi k_B T}.
\end{equation}
This equation is valid when the sticking coefficient of Au vapor impinging on the surface is near unity and there is no barrier layer on the surface that would suppress evaporation.

Data for $\ve$ for a sequence of pulses on a prepared sample is shown in Fig.~\ref{erosion velocity}b (black data). As we we have shown before \cite{Coppock2021,Coppock2022}, a plateau near $\ve=\ve(1337 \mathrm{K})=2.1\times10^{-10}$ m$\cdot \persec$  appears in the data when the particle reaches the Au melting point. The plateau arises because liquid Au is significantly less absorptive at 532 nm than solid Au\cite{Gerasimov2016}.  On the plateau, increasing or decreasing the pulse power increases or decreases the thickness of a thin layer of liquid Au on the surface, and the particle $T$ remains unchanged.

While reproducible data on the plateau can be obtained in a $p\simeq10^{-5}$ Torr \o2 atmosphere, the plateau occurs at a reduced $\ve$ once \o2 exposure is interrupted and the particle is subjected to repeated pulses in non \o2 ambient (Fig.~\ref{IntroPlot}b, red data). In order to investigate this suppression, a repeated sequence of pulses (Fig.~\ref{IntroPlot}a, inset) is used to sweep across the plateau.  When the flat region in the data is found, $\ve$ is determined and a plateau suppression factor $\sigma\equiv\ve/\ve(1337 \mathrm{K})$ is calculated.  Our working hypothesis is that $\sigma<1$ is an indication that an impermeable layer is present on the surface that impedes Au evaporation and leads to a measured $\ve$ that is less than the value predicted from Eqs. \ref{erosion velocity} and \ref{psat}. Full characterization of the behavior takes several hundred pulses that each typically consume 0.1\% of the mass of the particle. Because smaller samples reach higher temperatures for a given pulse power \cite{Coppock2022}, we scale the power (plotted in the inset to Fig.~\Ref{IntroPlot}a) by $(r/100~ \mathrm{nm})^2$ so that the plateau remains within the range of powers used in each sweep.

\section{Results and Discussion}
\subsection{Qualitative behavior in \o2 and vacuum ambient}
\label{sec:qualitative}

\begin{figure*}
    \centering
    \includegraphics[width=1.0\linewidth]{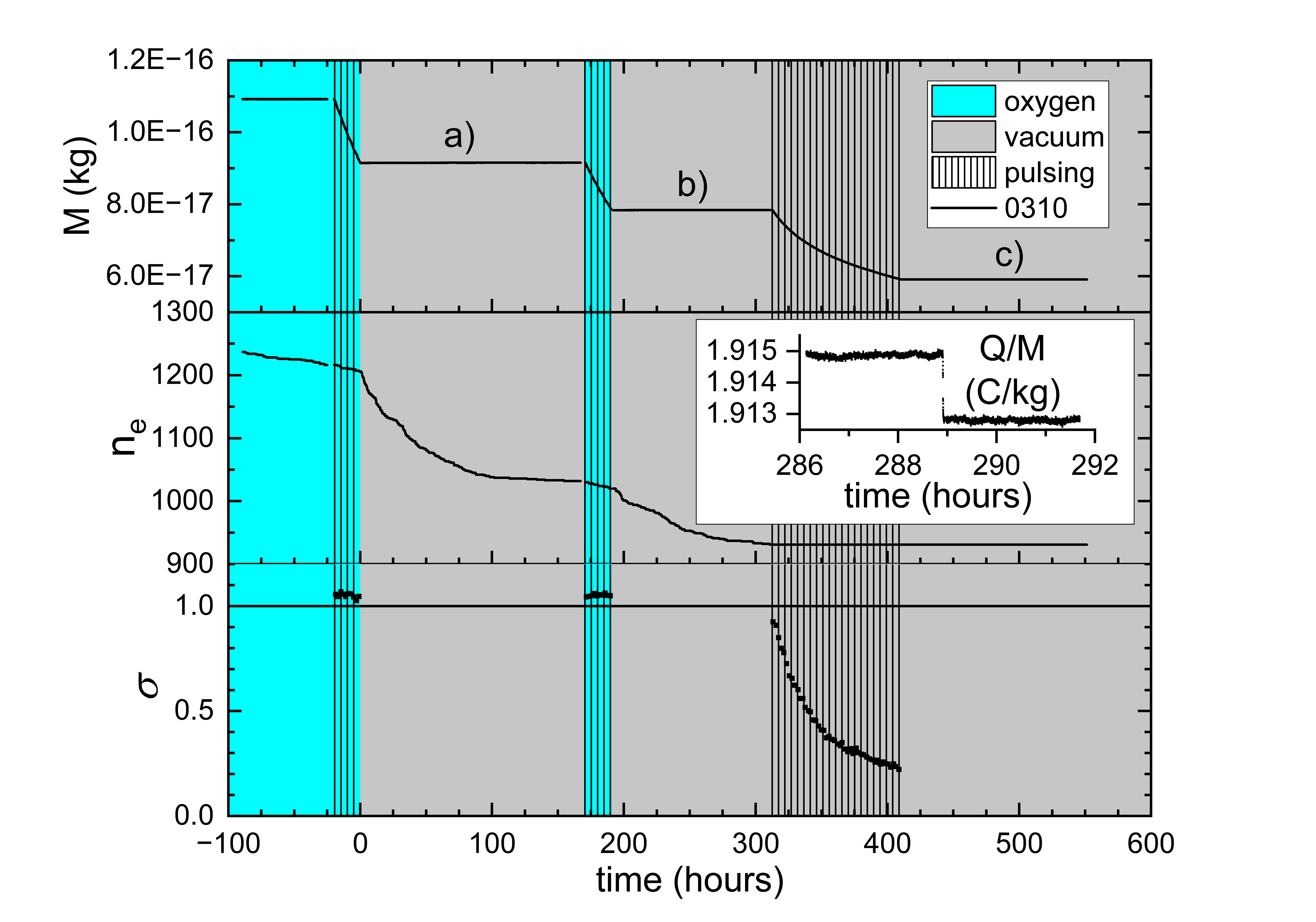}
    \caption{Behavior of an Au nanoparticle during a long series of pulses and quiescent periods.  Large mass loss occurs only during the heating pulses, but much smaller changes in mass also occur during the quiescent periods between pulses (see Fig.~\ref{Quiescent} for details on the plot segments marked ``a,b,c''). Charge loss accelerates after pulses in \o2, but diminishes (and even falls to zero) after long periods in vacuum.  The inset shows a drop in $\CMR$ associated with the loss of a single charge.  $\sigma\simeq$1 after preparation and cleaning in \o2. It is unaffected by a long duration quiescent period in vacuum but diminishes when the sample is exposed to pulses in vacuum.  
}
    \label{QualitativeFig}
\end{figure*}

For all data discussed in this paper, the initial pulse set is performed in \o2 at $p=2-3\times 10^{-5}~$ Torr, and consists of 120 laser pulses, separated by 10 minute intervals for data acquisition, in the sequence shown in the inset to Fig.~\ref{IntroPlot}a.  The sweep direction alternates between decreasing and increasing laser powers.  Subsequent pulse sets may be performed in vacuum or other gas ambients, and may have different time intervals between pulses.  Qualitative sample behavior in vacuum and \o2 ambient is shown in Fig.~\ref{QualitativeFig}.  The interval between pulses is 10 minutes throughout this data set.  Jumps in $\CMR$ during data acquisition occur in both the positive and negative direction.  Jumps due to charge loss lead to a well defined $decrease$ in $\CMR$ (Fig.~\ref{QualitativeFig}, inset) while $increases$ in $\CMR$ are caused by evaporation. Consequently, $M$ and $\ne$ can be separately tracked during data acquisition. Occasionally, charge loss events coincide with pulses, causing an error; however, since the plateau from which $\sigma$ is extracted generally contains five data points, outliers can be easily identified and discarded. Rarely, several data points on a plateau are marred by discharges, causing the calculated $\sigma$ data point to be an outlier.

The value of $\sigma$ derived from measurements of $\Delta M$ produced by the initial \o2 pulses consistently exceeds the expected value (=1) by about 5-10\%.  While  it is possible that this is a real effect (for example, a non-spherical particle would have $\sigma>1$ due to increased surface area), we cannot rule out that calibration errors in our measurement of $\CMR$ may also be responsible for this deviation.

After the last pulse in \o2 has been performed (which corresponds to $t$=0 for all our data), the \o2 leak valve is closed and the chamber pumps down in a few hours to $p<5\times10^{-8}$ Torr.  During a long quiescent period shown in Fig.~\ref{QualitativeFig} (when the particle is illuminated by the CW laser but is not subjected to laser pulses), the particle continues to discharge, but the discharge rate decelerates with time. After 175 hours the \o2 pressure is returned to its initial value, and the pulse sequence is repeated.  The long interval in vacuum has no effect on $\sigma$.

At $t=$ 195 hours the vacuum chamber is again pumped down to $p<5\times10^{-8}$ Torr.  The discharge rate increases after the pulses in \o2, but again slows down during the long quiescent interval.

Finally, at $t=$ 310 hours, a sequence of 576 pulses is performed (using the pattern in the inset to Fig.~\ref{IntroPlot} with a 10 minute interval between pulses) while the particle remains at $p<5\times10^{-8}$ Torr.  During this sequence of pulses, $\sigma$ drops from near unity to about 0.25.  During the pulses in vacuum, no discharge events occur, and none  occur during the 200 hour quiescent period subsequent to the vacuum pulses.

\subsection{Precision mass measurements}
\label{sec:precision}

\begin{figure}
%\begin{center}
\includegraphics[scale=1.0,draft=false]{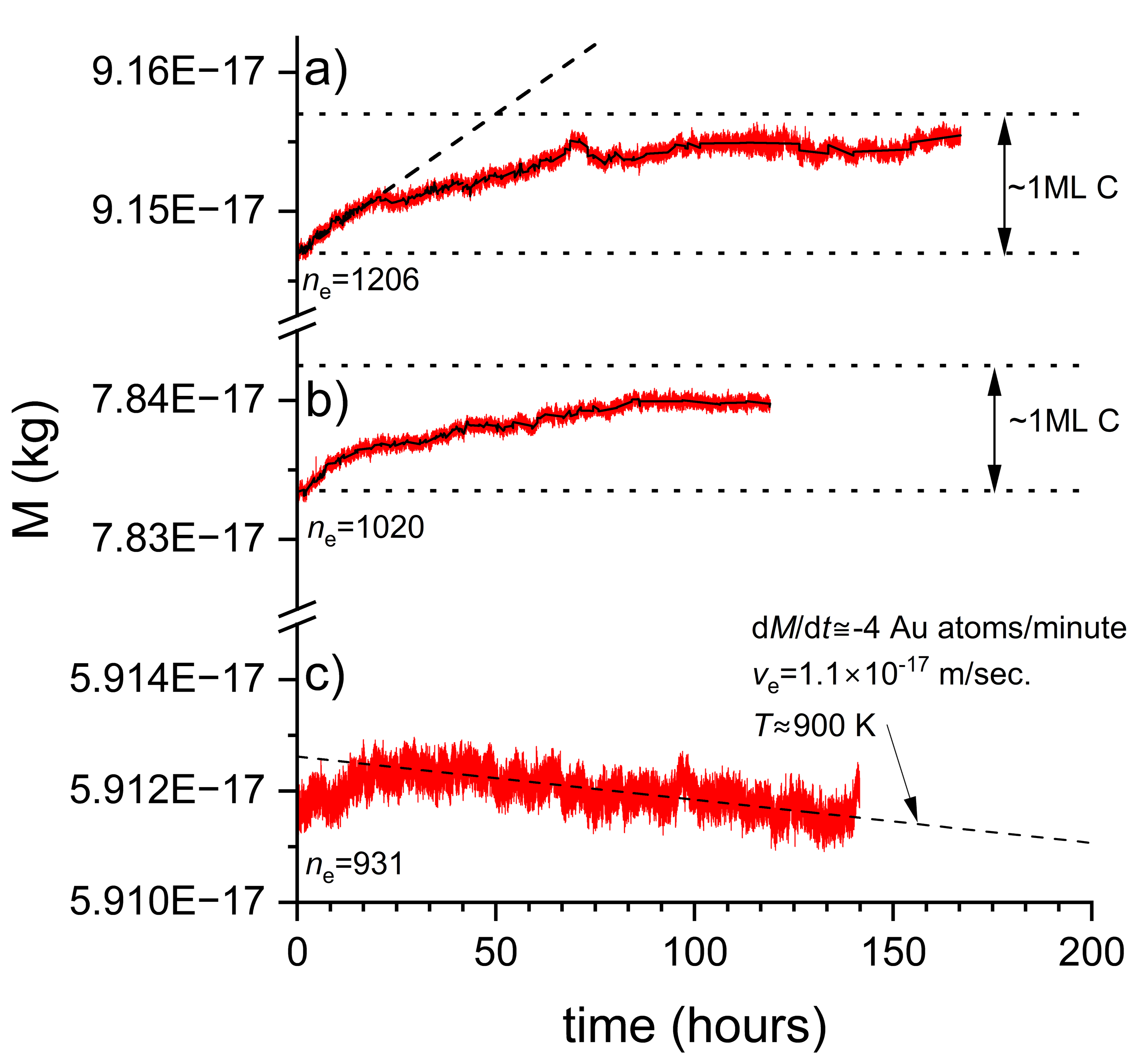} 
%\end{center}
\caption{
High resolution behavior of $M$ during the long quiescent periods marked ``a,b,c'' in Fig.~\ref{QualitativeFig}, with $t$=0 corresponding to the time of the last previous pulse.   Black solid lines are linear fits to the data between discharge events, while red lines are derived from $\CMR$ assuming the initial values of $\ne$ shown in the  figure, which is chosen so that the black and red data yield the same overall mass difference. During quiescent periods subsequent to pulsing in \o2, the particle $M$ increases and total $M$ change approaches the mass of a single monolayer of C ($7.6\times10^{-7}~$kg$~\permm$) surrounding an Au particle. After subjecting the sample to pulses in vacuum, $M$ decreases, consistent with Au evaporation at $T\simeq$900 K.  For the data in (c), no discharges occur, so data is obtained directly from Eq. \ref{Q/M} using $\ne$=931. 
}
\label{Quiescent}
\end{figure}

One clue as to the mechanism for the decline in $\sigma$ that we observe comes from more precise measurement of $M$ during the quiescent periods  between pulse sequences (Fig.~\ref{Quiescent}). Because discharge events are still occurring during quiescent periods, we obtain an estimate of the overall mass change during the period by fitting $\CMR$ between discharge events  to a line.  We then determine the initial $\ne$ by matching the mass difference obtained from  $\CMR$ and Eq. \ref{Q/M} to the overall mass difference obtained from the linear fits.

Using this technique, we observe that there is a small increase in $M$ in the quiescent periods subsequent to \o2 pulses (Figs.~\ref{Quiescent}a and \ref{Quiescent}b) and a smaller $M$ decrease after the pulses in vacuum (Fig.~\ref{Quiescent}c).  The maximum mass gain rate in Figs.~\ref{Quiescent}a and \ref{Quiescent}b is consistent with a monolayer of C accumulating on the surface in $\sim$50 hours, corresponding to a C-containing volatile gas in the chamber at $p=5\times10^{-11}$ Torr accumulating on the Au surface with unity sticking coefficient.  Both CO and \co2 are present in the chamber in sufficient quantities (Table \ref{TableRGATurbo}) to be the source of the mass increase that we observe.

The mass decrease (Fig.~\ref{Quiescent}c) is occurring on a surface with a barrier present ($\sigma<1$) that may prevent the accumulation of mass from impinging gases. The mass loss is consistent with Au evaporation from a surface with $\sigma$=0.25 at $T\sim$ 900 K, somewhat warmer than the $T$=750 K we have previously estimated \cite{Coppock2021} that was based on extrapolation of high temperature heating data at similar laser power densities.

\subsection{Effect of changes to the measurement procedure}
\label{sec:variations}

Assuming the validity of the hypothesis that the reduction in $\sigma$ is the consequence of the growth of a surface barrier layer, the simplest behavior would be that $\sigma$ would be purely a function of the time since measurements began in vacuum, since it is reasonable to assume that contaminants on the surface are accumulating at a constant rate.  Alternatively, since heating from the pulses is necessary for the formation of the surface barrier (Fig.~\Ref{QualitativeFig}), pulse heating could be the rate limiting step for barrier formation. Then, $\sigma$ would purely be a function of the  number of pulses in the contaminated environment. 

\begin{figure}
%\begin{center}
\includegraphics[scale=1.0,draft=false]{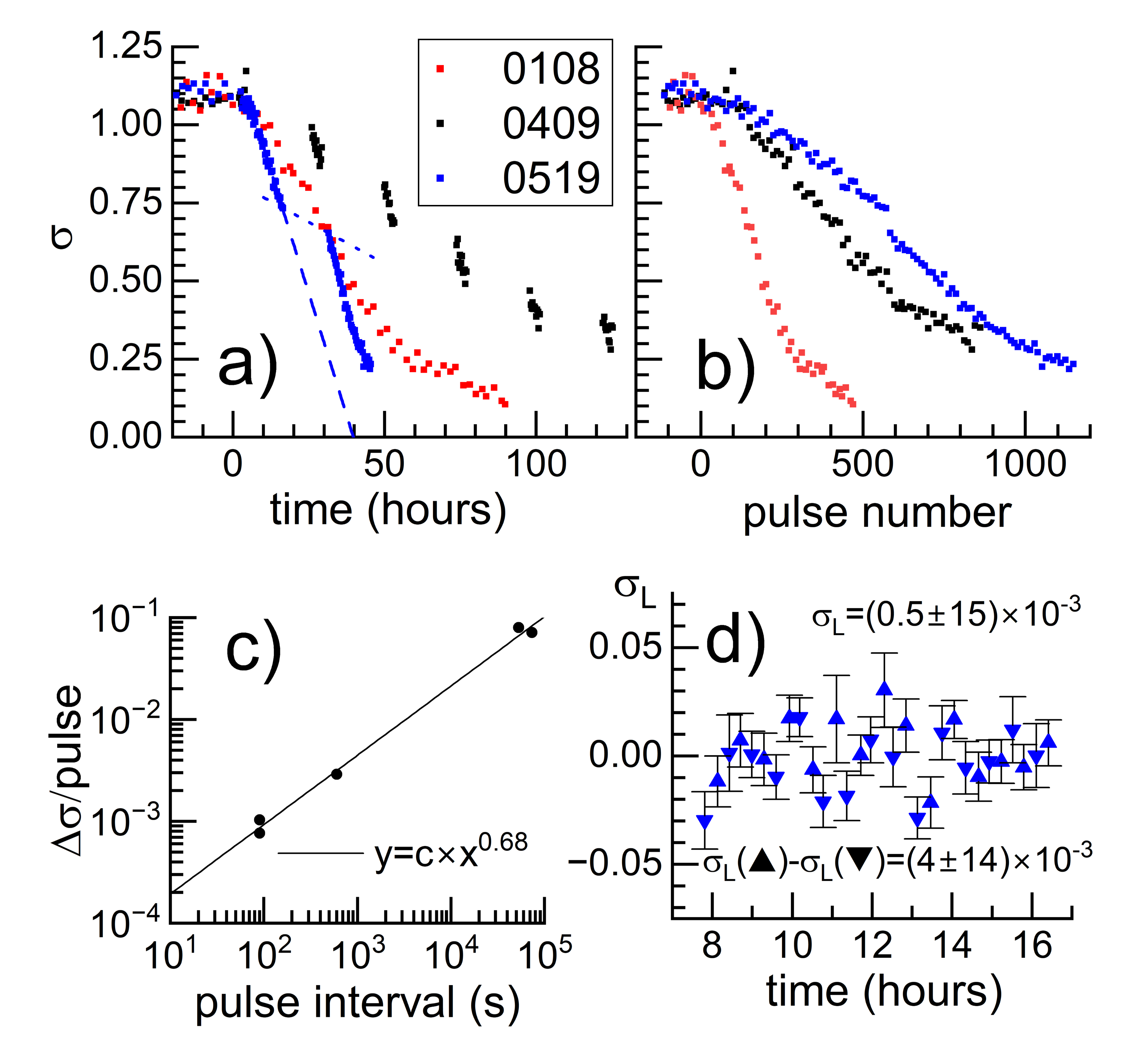} 
%\end{center}
\caption{
$\sigma$ data for three samples plotted as a function of time (a) and pulse number (b). For Sample 0108, data was taken using pulses spaced at at a constant 600 second interval, while for 0409 and 0519 data were taken using separate sets of pulses spaced at 90 second intervals. (c) Dependence of $\Delta \sigma$ on the interval $\Delta t$ between pulses.  For intervals <1000 seconds, data was taken using a linear fit to a sequence of pulses (dashed line in (a)), while  data at larger intervals was taken using two points before and after a long period of pulse cessation (dotted line in (a)). (d) Selected data for 0519, replotted as $\sigma_L$ after subtracting a linear fit. $\blacktriangle$ and $\blacktriangledown$ correspond respectively to data taken when sample temperature is either being swept upwards or downwards.
}
\label{PulseRateDirection}
\end{figure}

To test these ideas, we performed experiments on samples with different pulse intervals (600 s and 90 s) and plotted the data, both as a function of $t$ and of pulse number (Figs.~\ref{PulseRateDirection}a and \ref{PulseRateDirection}b).  The data for $\sigma$  are consistent with neither pure time or pulse number dependence. The rough dependence of $\Delta \sigma$/pulse on pulse interval $\Delta t$ is plotted in Fig.~\ref{PulseRateDirection}c.  The dependence is sublinear $\Delta \sigma \propto \Delta t^{0.68}$ but continues to increase  for $\Delta t$< 1 day.  This behavior suggests that contaminants can accumulate, albeit at a decreasing rate, on the surface between pulses, but are incorporated into the barrier layer only after pulsing. 

Since we have established pulsing is necessary for the formation of the barrier layer, another issue to address is whether the power of the pulses (and the consequent peak $T$ the particle experiences) has an impact on the rate of barrier formation.  Data taken on a plateau on a downward sweep of pulse power occurs immediately after the sample has been exposed to $T$ above the Au melting point, while data on an upward sweep is taken immediately after it is pulsed at $T$ below the melting point.  To determine if these data sweeps differ, we have fit $\sigma$ data in a linear region (along the dashed line in Fig.~\ref{PulseRateDirection}a) and plot the residual $\sigma_L$ for both upward and downward sweeps (Fig.~\ref{PulseRateDirection}d).  The data indicates that the downward sweeps (preceded by high temperature exposure) cause a slightly greater reduction in $\sigma$ than upward sweeps, but the effect is significantly smaller than statistical variation in the data.  We conclude that raising $T$ above the Au melting point is not necessary and does not strongly promote barrier formation on the particle.

\subsection{Re-exposure to oxygen}
\label{sec:oxygen}

\begin{figure}
%\begin{center}
\includegraphics[scale=1,draft=false]{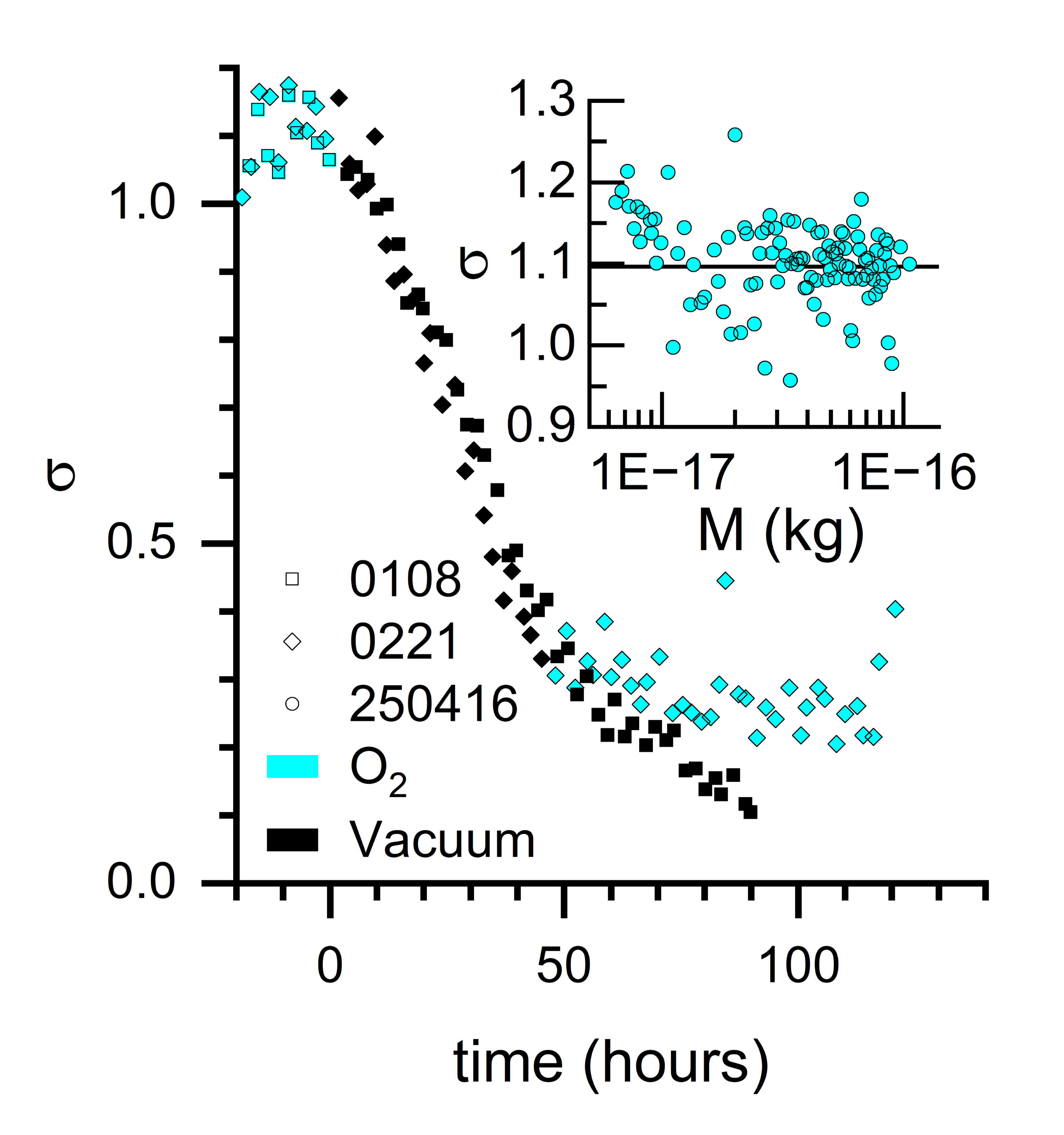} 
%\end{center}
\caption{
$\sigma$, plotted against time for Samples 0108 and 0221 using constant 600 second pulse interval.  Data for both samples is in vacuum for $0\leq t \leq$ 48 hours, but 0221 is re-exposed to \o2 at $t$>48 hours. For $t$>48 hours, $\sigma$ continues to decline for the sample in vacuum but remains roughly constant for the sample in \o2.   Inset shows data for a sample (250416) that was continuously exposed to \o2 as it was gradually evaporated by repeated pulses.  For this sample, $\sigma$ varies little from the nominal value of $\simeq$1.1 that we have observed in all samples.
}
\label{OxygenPlot}
\end{figure}

Since we use \o2 to prepare and clean the Au nanoparticle surface prior to data collection, we performed experiments to determine if the surface could be recleaned (and the hypothetical barrier removed) by subsequent re-exposure to \o2.  Data is presented for two samples in Fig.~\ref{OxygenPlot}. After standard preparation in \o2, both are pulsed in vacuum for 48 hours and show nearly identical $\sigma$ reduction with time.  At $t=$ 48 hours, one sample is re-exposed to \o2.  While the sample that remains in vacuum continues to exhibit $\sigma$ reduction, the sample in \o2 maintains an approximately constant $\sigma$.  The \o2-treated sample does not recover a clean surface $\sigma \geq1$.  These data suggest that the peak $T$ the samples experience during pulsing ($\geq$1350 K) are insufficient to oxidize and remove the surface barrier on the sample in an \o2 ambient. 

In the inset to Fig.~\ref{OxygenPlot}, data is shown for a sample that was subjected to repeated pulses in \o2 until $M<7\times10^{-18}$ kg.  Even though $M$ changes by more than an order of magnitude, $\sigma$ remains nearly constant at $\simeq$1.1. Thus, the continued suppression of $\sigma$ after Sample 0221 is re-exposed to \o2, is not a  consequence of the  particle's diminishing mass, which was $>6\times10^{-17}$ at then end of data collection.

\subsection{Sample exposure to other gases}
\label{sec:gases}

To determine the source of the carbon that we hypothesize is accumulating on Au surfaces, we performed a series of experiments where Au nanoparticles are exposed to various gases at $p\simeq6\times10^{-6}$ Torr, about two orders of magnitude greater than the total pressure in our experiments in nominal vacuum. Carbon containing gases CO, \c2h4, \co2, as well as \H2 and He were tested. We observed no significant acceleration of the decline in $\sigma$ in any of these tests. This result strongly implies either that we have not yet identified the source of the C in our experiments or that the partial pressure of the C source is not the rate limiting factor in the growth time of the barrier layer on Au surfaces.

\begin{figure}
%\begin{center}
\includegraphics[scale=1.0,draft=false]{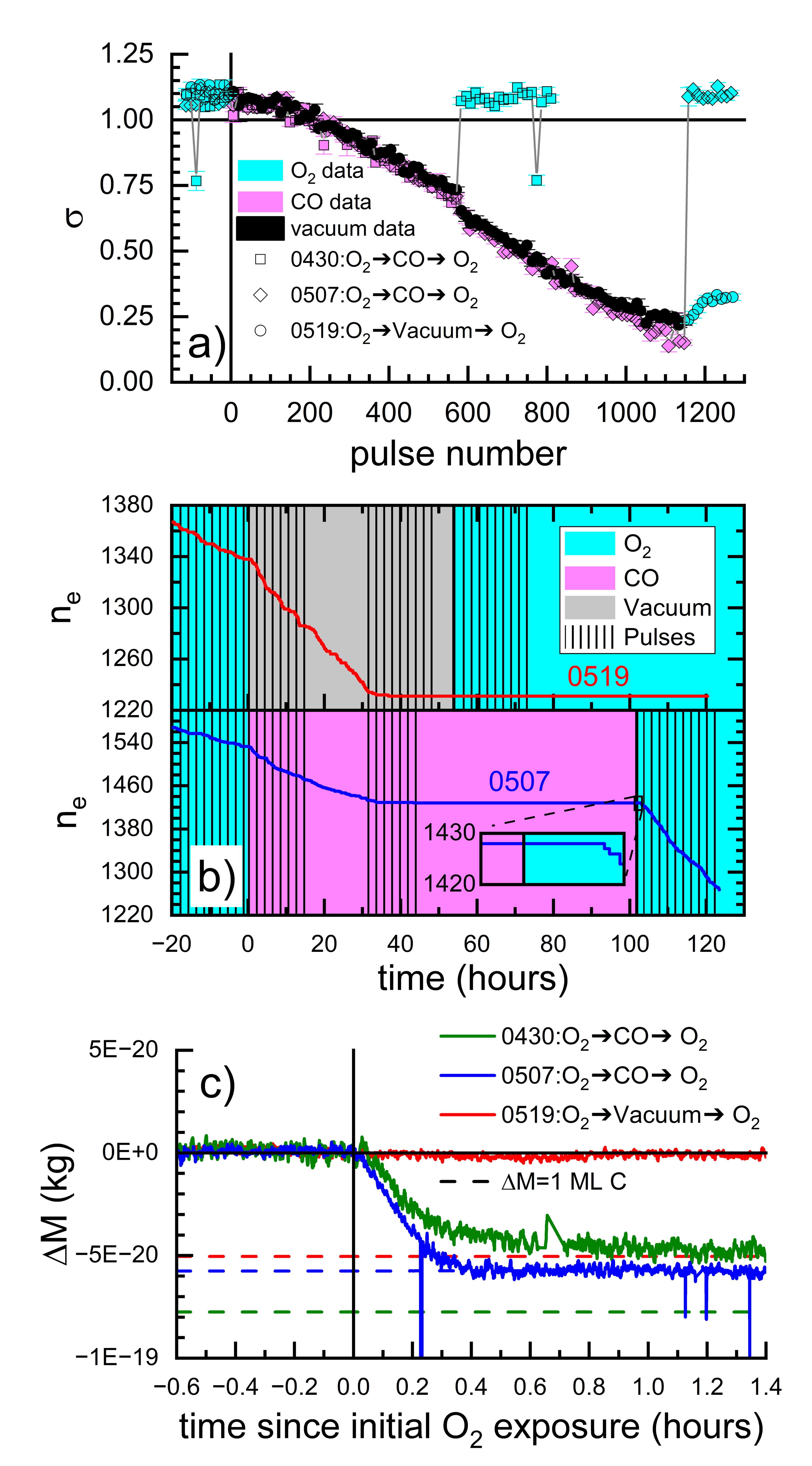} 
%\end{center}
\caption{
(a) Effect of re-exposure to \o2 for samples initially pulsed in CO (0430 and 0507) and in vacuum (0519):  for 0430, the sample was subjected to a sequence of 576 pulses at 90 second intervals in CO ambient prior to switching to \o2 ambient and initiating a pulse sequence at 600 second intervals. The process for 0507 was the same as 0430, except that the sample was subjected to two sets of rapid pulses in CO. For both samples, $\sigma \rightarrow >1$ immediately after \o2 is reintroduced. The pulse sequence for 0519 is identical to that of for 0507, but it was in a vacuum environment prior to \o2 re-exposure. For this sample, $\sigma$ increases only slightly when \o2 is reintroduced.  (b) Discharge behavior for sample in vacuum and sample exposed to CO:  for both  samples, discharge ceases when $\sigma\simeq$0.75.  Discharge resumes, however, when the CO sample is re-exposed to \o2.  Discharge resumes prior to initiation of pulsing in \o2. Time range in magnified region is the same as that plotted in (c). (c) Mass data for the samples when they are re-exposed to \o2: $t=0$ corresponds to when $p>10^{-5}$ Torr.  After \o2 exposure, $M$ for the samples previously exposed to CO lose a mass comparable to that of a monolayer of C on their surface.  No similar effect is seen for the sample previously in a vacuum environment.  
}
\label{MassBumpPlot}
\end{figure}

While the rate of decline of $\sigma$ shows little dependence on the presence of added gases we have tested,  the behavior of $\sigma$ when samples are re-exposed to \o2 differs dramatically between samples exposed to CO and those samples treated in vacuum.  This behavior is depicted in Fig.~\ref{MassBumpPlot}a:  after preparation in \o2, all three samples shown were subjected to sequences of rapid (interval=90 s) pulses.  Sample 0430 was subjected to a single set of 576 pulses, and Sample 0507 was subjected to two sets of pulses, both in CO ambient.  Sample 0519 was subjected to the same two pulse set sequence as Sample 0507, but was pulsed in vacuum.  After these pulse sequences, all samples were re-exposed to \o2 and subsequently subjected to a sequence of pulses (interval=600 s) in the \o2 ambient.

For both samples exposed to CO, $\sigma$, determined from the first pulse data in \o2, completely recovers to its pristine surface value (1.1). The sample treated in vacuum shows a small increase of $\sigma$ after exposure to \o2, but soon stabilizes at a value comparable to its minimum value after vacuum pulse  treatment.

Changes in behavior of the CO-exposed samples upon \o2 re-exposure occur even prior to the initiation of the pulse sequence in \o2. Fig.~\ref{MassBumpPlot}b shows that charge loss of the two samples subjected to two sequences of pulses prior to re-exposure to \o2. Both the vacuum sample and the CO sample cease their discharging when $\sigma$ falls below $\sim$0.7. The vacuum-treated sample completely stops discharging after $t$=37 hours, and does not resume discharge after re-exposure to \o2 and pulsing.  The CO-treated sample stops discharging at $t$=43 hours but resumes discharging when it is re-exposed to \o2.  The magnified region shows that discharging resumes even prior to the first pulses in \o2.

For the samples treated in CO, re-exposure to \o2 results in a rapid mass loss of the particle (Fig.~\ref{MassBumpPlot}c).  Mass loss begins as soon as \o2 is admitted into the chamber and reaches a new constant value within 30 minutes.  For the sample exposed to two pulse sequences (0507) in CO  that attained a minimum $\sigma$=0.15, the change in mass, $\Delta M$, is very close to that of a monolayer of C on a spherical surface with radius given by the particle's mass and the Au density. The CO-exposed sample subjected to only a single pulse sequence had  $\Delta M$ equivalent to 2/3 of a monolayer of C on its surface. The sample pulsed in vacuum and re-exposed to \o2 had no discernible $\Delta M$.

We have also measured the effect of \o2 re-exposure on samples pulsed  in He and have found similar behavior to samples treated in vacuum ($\sigma$ does not significantly increase upon \o2 re-exposure) . We surmise that exposure to CO creates a fragile barrier layer on the Au surface that is oxidized and removed at the temperature of the sample under  CW laser illumination ($\leq$ 900 K).

We performed numerous other experiments in CO ambient to elucidate the mechanism for the decline of $\sigma$ we observe. We doubled (to $p=1.5\times10^{-5}$ Torr) and quadrupled the CO pressure and saw no pressure effect. Doubling the CW laser intensity between pulses had no discernible effect on $\sigma$ behavior, and turning off ion sources from ion gauges also had negligible effect. Only changing the pulse interval of the pulse pattern affected the $\sigma$ decline: larger pulse intervals led to larger $\Delta \sigma$ per pulse, consistent with data taken in vacuum discussed in Section \ref{sec:variations}.

\subsection{Possible mechanisms of barrier formation}
\label{sec:ideas}

In our experiments, we have not been able to establish definitively that CO is the source material for barrier formation, or even that C is forming the barrier.  We believe  that observation of $\sim1$ monolayer C mass gained and lost during our measurements (Figs.~\ref{Quiescent} and \ref{MassBumpPlot}c) is good evidence that the barrier to Au evaporation that we observe is a graphene layer.  That the mass loss behavior is only observed in CO  ambient suggests that CO is relevant to the formation of the barrier.

\begin{figure}
%\begin{center}
\includegraphics[scale=1.0,draft=false]{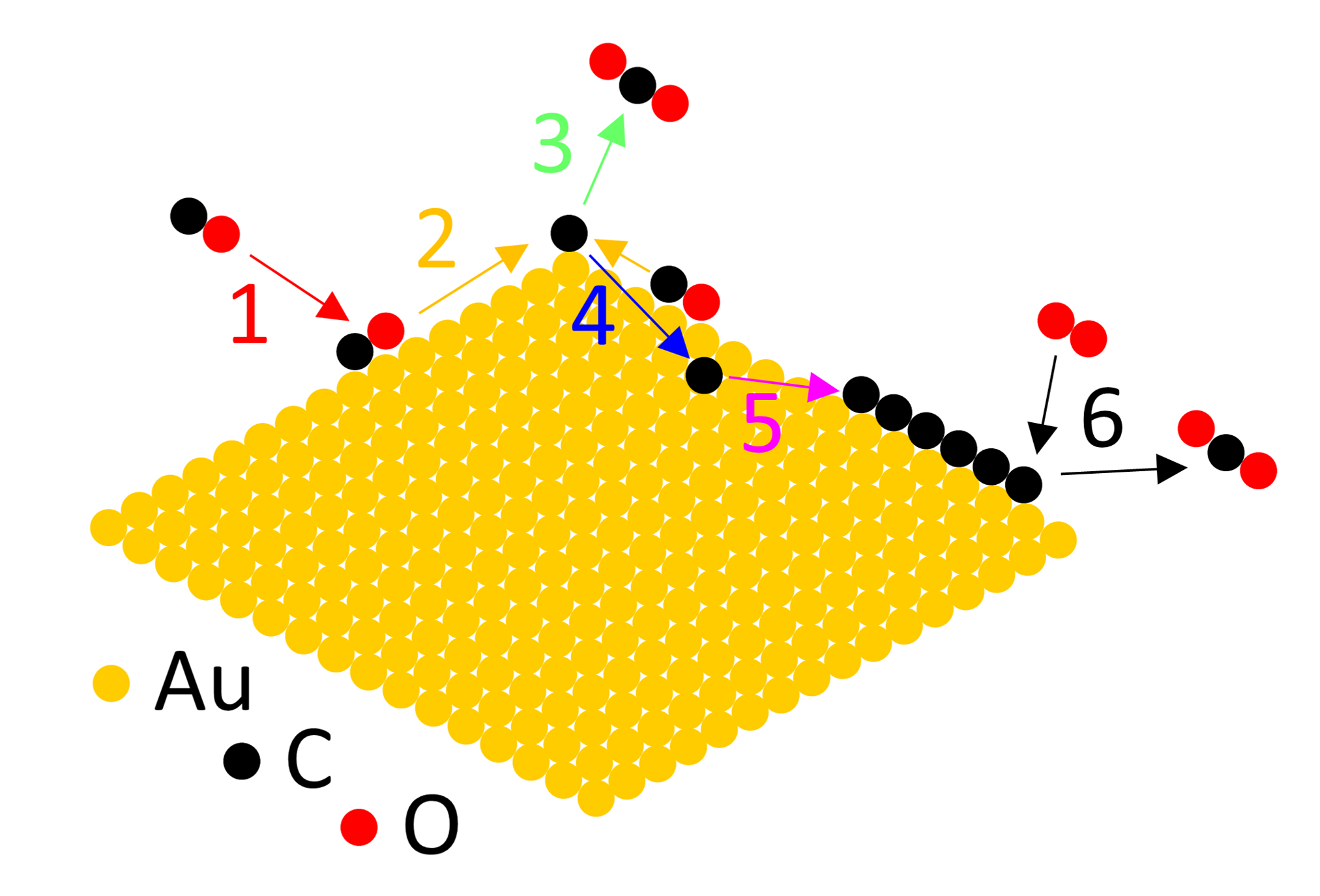} 
%\end{center}
\caption{
Proposed reaction pathway for formation of a C barrier layer on a levitated Au nanoparticle. (1) CO (or possibly another C-containing gas) is adsorbed on the Au surface into a mobile, weakly bonded state. (2) Two CO molecules diffuse to a site (here depicted as a vertex of a 2D crystal) where the reaction 2CO$\rightarrow$C+\co2 proceeds (3). The C is strongly attached to the reaction site as long as the nanoparticle is solid and its presence prevents further reactions at the site. (4) When the particle is heated or melted by a laser pulse, the C can leave the site and move to a new mobile location, either inside the particle or strongly bonded to its surface. This step both frees the vertex site for a new reaction and allows the particle to migrate and bond to a graphene fragment on the Au surface (5).   Finally, in the presence of \o2, the graphene barrier layer can be  oxidized and removed if the reaction kinetics are favorable (6).
}
\label{IdeasFigure}
\end{figure}

Based on these assumptions, we propose the reactions portrayed in Fig.~\ref{IdeasFigure} are responsible for our observations: during the intervals between pulses, CO is adsorbed onto the Au surface and diffuses to reaction sites that facilitate the reaction: 2CO$\rightarrow$ \co2+C. Because the sample is always illuminated with 532 nm laser radiation between pulses, plasmon resonance may significantly accelerate this reaction. Once the reaction has occurred however, the presence of C on this site prevents further reactions, and it is necessary to either heat or melt the sample to allow the C to leave the site or to create new sites (by cooling and recrystallization) for new reactions to occur.

Pulse heating and melting of the sample allows the  reacted C to diffuse to graphene-like layers, 
which presumably nucleate on the surface when enough C is present.  Long repetitions of pulse sequences, like those that we apply to the sample, lead to the growth of the surface graphene layer that ultimately forms the barrier to Au evaporation  that we observe.

In the presence of \o2, C in the Au particle or on its surface can be removed by oxidation reactions.  To explain why the barrier layers formed in CO ambient are removed at low temperatures,  while those that form in vacuum conditions are stable at much higher temperatures, we speculate that graphene layers grown with high concentrations of CO have zigzag edges, while the graphene layers grown in vacuum (or much lower CO pressure) have armchair terminations that are  much less readily oxidized than zigzag edges \cite{Xu2014}.

During the interval between pulses, the particle is solid and is presumably crystallized in the face centered cubic (fcc) structure.  Edges and corners of the crystal are natural candidates for the special sites we hypothesize are necessary for the barrier formation\cite{Barmparis2012}.  Reactions at these sites are the rate limiting step for barrier growth and are the reason that changes in CO pressure do not have a significant effect on the rate of barrier growth. It is likely that slow C diffusion away from the reaction sites can occur without high temperature pulsing, allowing reactions at the site to proceed at diminishing rates over long periods. This diffusion would explain both the mass changes we observe in Fig.~\ref{QualitativeFig} in the periods when no pulses were applied and the sublinear increase in $\Delta \sigma$ that we see with long pulses intervals in \ref{PulseRateDirection}c.

\subsection{Particle discharging}
\label{sec:discharge}

It is unlikely that the surface reactions that lead to loss of charge on the particle are necessary for the barrier growth, since we typically lose only a few hundred charges during a full experimental pulse sequence (Table \ref{TableMassCharge}).  As previously noted, discharge is fastest on a sample pulsed in \o2 ambient and shows no sign of diminishing over time. In ambients other than \o2, discharge ceases after long exposures ($\sim$days), or when $\sigma\leq$ 0.7.

Because all of our samples are positively charged, discharge can either occur when a positive ion (like \h+ or \Au+)  leaves the particle or a negative ion (or \e-) becomes attached to it.  We believe that discharging from incoming negative charges is unlikely, because a conducting graphene surface layer should not be an impediment to charge, and since we observe that discharging ceases entirely when the barrier layer only partially covers the surface.

Our experiments in \H2 ambient were intended to test whether \H2 reactions on the Au surface \cite{Watkins2017,Mukherjee2012,Mukherjee2014,Kar2025} were relevant, either to discharging or barrier formation.  We found that the presence of \H2 did not affect the  behavior of $\sigma$ or the discharge rate significantly.  Also in \H2 ambient, we completely blocked the CW laser for 24 hours subsequent to our usual \o2 pulse pre-treatment.  By measuring $\CMR$ before and after the light interruption, we established that particle discharge continued at usual rates, even in the complete absence of 532 nm excitation. While we have no adequate explanation of the particle discharge behavior we have observed, we hypothesize that rare surface states similar to those responsible for C uptake are playing a role in charge reactions; these reactions may also be hindered by accumulating C on the surface of the Au nanoparticle.

\section{Conclusions}
\label{sec:conclusions}

We have performed experiments to probe the barrier to evaporation that forms in non-\o2 ambients when levitated Au nanoparticles are repeatedly heated to the Au melting point by  532 nm laser pulses. We hypothesize that this barrier is a graphene layer that grows on the Au surface after exposure to C-containing gases and repeated heating and melting of the particle.  This hypothesis is supported by our observation that a mass comparable to that of a graphene monolayer accumulates on the surface over time, and a similar amount of mass is lost when samples are re-exposed to \o2 after pulses in a CO ambient.

Because photocatalyzed CO reactions on Au surfaces are well documented \cite{Hung2008,Hung2010} and  because it is a significant residual gas in our vacuum chamber, we believe that CO is the most likely source for the C arriving at the particle.  Because we do not observe a change of the rate of accumulation of the barrier when we add CO, we suggest that bottlenecks associated with surface states that catalyze the 2CO$\rightarrow$C+\co2 reaction are what determines the rate of growth of the graphene barrier layer, rather than CO pressure.

Further experiments will be necessary to more fully elucidate the mechanism for the formation of the barrier that is responsible for our observations.  Most obviously, experiments will need to be performed in a better vacuum so that measurements can be carried out in the regime where the layer growth rate does depend on the pressure of added gases. Recently, measurements on nanoscale objects levitated in ion traps have been performed in the ultra-high-vacuum (UHV) pressure regime ($p=5\times10^{-11}$ Torr)\cite{Dania2024}, and we expect that our current measurement techniques will be fully adaptable to a UHV environment\cite{Kane2026}. 

Also, because the critical reactions necessary to form the barrier are likely photocatalytic in nature, more experiments are necessary to determine the effect of varying the wavelength and intensity---during and between pulses---of the laser irradiation. Completely interrupting light exposure for short periods may help to determine if radiation is necessary for the reactions to proceed. While these latter experiments will be challenging--since illumination is necessary to maintain the  stability of the confined particle in high vacuum environments--short term interruptions should be possible in traps with minimal excess noise\cite{Kane2026}. 

Recently, we have shown that levitated particles can be deposited on a nearby substrate for $ex~situ$ analysis\cite{Coppock2024}.  This technique may prove helpful for detailed compositional analysis of the Au nanoparticle, both before and after the barrier has formed.
Finally, if our hypothesis is correct that graphene layers are forming on levitated Au nanoparticle surfaces, measurements on levitated Au-graphene bilayers\cite{Kim2026}--samples that have a graphene seed layer present from the outset--may provide new insights on Au-catalyzed graphene crystal growth of levitated materials.


\section*{Data Availability}
Data relevant to this study are available from the authors upon reasonable request.

\begin{acknowledgments}
This work was supported by the Laboratory for Physical Sciences, Contract \#H9823023C0086.
\end{acknowledgments}

\section*{Author Contributions}
\textbf{J.E.C.:} Methodology, Investigation, Writing -- original draft, Writing -- review \& editing. \textbf{S.K.:} Investigation, Writing -- review \& editing. \textbf{B.E.K.:} Conceptualization, Software, Methodology, Visualization, Writing -- original draft; Writing -- review \& editing.

\section*{Notes}
The authors declare no competing financial interest.

\bibliography{Encapsulation}

\end{document}